\documentclass[11pt]{article}

\usepackage{amsmath, amssymb, amsthm}

\newtheorem{theorem}{Theorem}
\newtheorem{lemma}{Lemma}
\newtheorem{corollary}{Corollary}
\newtheorem{proposition}{Proposition}

\newcommand{\N}{\mathbb{N}}

\newcommand{\B}{\mathbb{B}}
\newcommand{\Per}{\mathcal{P}}
\newcommand{\AP}{\mathcal{AP}}
\newcommand{\MT}[1]{\mathop{\mathrm{MT}\langle \N, <, #1\rangle}} % monadic theory
\newcommand{\poly}{\mathop{\mathrm{poly}}}
\title{On Almost Periodicity Criteria for Morphic Sequences\\
in Some Particular Cases}

\author{Yuri Pritykin%
\thanks{Moscow State University, Russia,
\texttt{http://lpcs.math.msu.su/\~{}pritykin/},
\texttt{yura@mccme.ru}. The work was partially supported by RFBR
grants 06-01-00122, 05-01-02803, Kolmogorov grant of Institute of
New Technologies, and August M\"obius grant of Independent
University of Moscow.}}

\begin{document}

\maketitle

\begin{abstract}
In some particular cases we give criteria for morphic sequences to
be almost periodic (=uniformly recurrent). Namely, we deal with
fixed points of non-erasing morphisms and with automatic sequences.
In both cases a polynomial-time algorithm solving the problem is
found. A result more or less supporting the conjecture of
decidability of the general problem is given.
\end{abstract}

\section{Introduction} \label{Introduction}

Different problems of decidability in combinatorics on words are
always of great interest and difficulty. Here we deal with two main
types of symbolic infinite sequences~--- morphic and almost
periodic~--- and try to understand connections between them. Namely,
we are trying to find an algorithmic criterion which given a morphic
sequence decides whether it is almost periodic.

Though the main problem still remains open, we propose
polynomial-time algorithms solving the problem in two important
particular cases: for pure morphic sequences generated by
non-erasing morphisms (Section~\ref{Pure}) and for automatic
sequences (Section~\ref{Uniform}). In Section~\ref{Monadic} we say a
few words about connections with monadic logics. In particular, in a
curious result of Corollary~\ref{ap-morphic-regulator-corollary} we
give a reason why the main problem may be decidable.

Some attempts to solve the problem were already done.
In~\cite{Cobh72} A.~Cobham gives a criterion for automatic sequence
to be almost periodic. But even if his criterion gives some
effective procedure solving the problem (which is not clear from his
result, and he does not care about it at all), this procedure could
not be fast. We construct a polynomial-time algorithm solving the
problem. In \cite{Maes00} A.~Maes deals with pure morphic sequences
and finds a criterion for them to belong to a slightly different
class of generalized almost periodic sequences (but he calls them
almost periodic~--- see \cite{Prit06c} for different definitions).
And again, his algorithm does not seem to be polynomial-time.

All the results of this paper can be found in~\cite{Prit07a}.

\section{Preliminaries} \label{Preliminaries}

Denote the set of natural numbers $\{0, 1, 2,\dots\}$ by $\N$ and
the binary alphabet $\{0, 1\}$ by $\B$. Let $A$ be a finite
alphabet. We deal with sequences over this alphabet, i.~e., mappings
$x\colon \N \to A$, and denote the set of these sequences by~$A^\N$.

Denote by $A^*$ the set of all finite words over $A$ including the
empty word~$\Lambda$. If $i\le j$ are natural, denote by $[i,j]$ the
segment of $\N$ with ends in~$i$ and~$j$, i.~e., the set $\{i, i+1,
i+2,\dots,j\}$. Also denote by $x[i,j]$ a subword $x(i)x(i+1)\dots
x(j)$ of a sequence $x$. A segment $[i,j]$ is an occurrence of a
word $u\in A^*$ in a sequence $x$ if $x[i,j] = u$. We say that $u
\ne \Lambda$ is a factor of $x$ if $u$ occurs in $x$. A word of the
form $x[0,i]$ for some $i$ is called prefix of $x$, and respectively
a sequence of the form $x(i)x(i+1)x(i+2)\dots$ for some $i$ is
called suffix of $x$ and is denoted by $x[i,\infty)$. Denote by
$|u|$ the length of a word~$u$. The occurrence $u = x[i,j]$ in $x$
is $k$-aligned if $k|i$.

A sequence $x$ is periodic if for some $T$ we have $x(i) = x(i + T)$
for each $i\in\N$. This $T$ is called a period of $x$. We denote
by~$\Per$ the class of all periodic sequences. Let us consider an
extension of this class.

A sequence $x$ is called \emph{almost periodic%
\footnote{It was called \emph{strongly} or \emph{strictly almost
periodic} in~\cite{MuchSemUsh03,Prit06b}.}%
} if for every factor~$u$ of $x$ there exists a number~$l$ such that
every factor of $x$ of length~$l$ contains at least one occurrence
of~$u$ (and therefore $u$ occurs in $x$ infinitely many times).
Obviously, to show almost periodicity of a sequence it is sufficient
to check the mentioned condition only for all prefixes but not for
all factors (and even for some increasing sequence of prefixes
only). Denote by $\AP$ the class of all almost periodic sequences.

Let $A$, $B$ be finite alphabets. A mapping $\phi\colon A^*\to B^*$
is called \emph{a morphism} if $\phi(uv) = \phi(u)\phi(v)$ for all
$u,v\in A^*$. A morphism is obviously determined by its values on
single-letter words. A morphism is \emph{non-erasing} if $|\phi(a)|
\geqslant 1$ for each $a \in A$. A morphism is $k$-uniform if
$|\phi(a)| = k$ for each $a \in A$. A 1-uniform morphism is called a
coding. For $x \in A^\N$ denote
  $$
 \phi(x) = \phi(x(0))\phi(x(1))\phi(x(2))\dots
  $$
Further we consider only morphisms of the form $A^* \to A^*$ (but
codings are of the form $A \to B$, which in fact does not matter,
they can be also of the form $A \to A$ without loss of generality).
Let $\phi(s) = su$ for some $s\in A$, $u\in A^*$. Then for all
natural $m < n$ the word $\phi^n(s)$ begins with the word
$\phi^m(s)$, so $\phi^\infty(s) = \lim_{n \to \infty} \phi^n(s) =
su\phi(u)\phi^2(u)\phi^3(u)\dots$ is well-defined. If $\forall n\
\phi^n(u) \ne \Lambda$, then $\phi^\infty(s)$ is infinite. In this
case we say that $\phi$ is \emph{prolongable} on $s$. Sequences of
the form $h(\phi^\infty(s))$ for a coding $h\colon A \to B$ are
called \emph{morphic}, of the form $\phi^\infty(a)$ are called
\emph{pure morphic}.

Notice that there exist almost periodic sequences that are not
morphic (in fact, the set of almost periodic sequences has
cardinality continuum, while the set of morphic sequences is
obviously countable), as well as there exist morphic sequences that
are not almost periodic (you will find examples later). Our goal is
to determine whether a morphic sequence is almost periodic or not
given its constructive definition.

First of all, observe the following

\begin{lemma} \label{first-criterion}
A sequence $\phi^\infty(s)$ is almost periodic iff $s$ occurs in
this sequence infinitely many times with bounded distances.
\end{lemma}
\begin{proof}
In one direction the statement is obviously true by definition.

Suppose now that $s$ occurs in $\phi^\infty(s)$ infinitely many
times with bounded distances. Then for every $m$ the word
$\phi^m(s)$ also occurs in $\phi^\infty(s)$ infinitely many times
with bounded distances. But every word $u$ occurring in
$\phi^\infty(s)$ occurs in some prefix $\phi^m(s)$ and thus occurs
infinitely many times with bounded distances.
\end{proof}

For a morphism $\phi\colon \{1,\dots,n\} \to \{1,\dots,n\}$ we can
define a corresponding matrix $M(\phi)$, such that $M(\phi)_{ij}$ is
a number of occurrences of symbol $i$ into $\phi(j)$. One can easily
check that for each $l$ we have $M(\phi)^l = M(\phi^l)$.

Morphism $\phi$ is called \emph{primitive} if for some $l$ all the
numbers in $M(\phi^l)$ are positive.

Let us construct an oriented graph $G$ corresponding to a morphism.
Let its set of vertices be $A$. In $G$ edges go from $b \in A$ to
all the symbols occurring in~$\phi(b)$.

For $\phi^\infty(s)$ it can easily be found using the graph
corresponding to $\phi$ which symbols from $A$ really occur in this
sequence. Indeed, these symbols form the set of all vertices that
can be reached from~$s$. So without loss of generality from now on
we assume that all the symbols from $A$ occur in $\phi^\infty(s)$.

A morphism is primitive if and only if its corresponding graph is
strongly connected, i.~e., there exists an oriented path between
every two vertices. This reformulation of the primitiveness notion
seems to be more appropriate for computational needs.

By Lemma~\ref{first-criterion} (and the observation that codings
preserve almost periodicity) morphic sequences obtained by primitive
morphisms are always almost periodic. Moreover, in the case of
increasing morphisms (such that $|\phi(b)| \geqslant 2$ for
each~$b$) this sufficient condition is also necessary (and this is a
polynomial-time algorithmic criterion). However when we generalize
this case even on non-erasing morphisms, it is not enough to
consider only the corresponding graph or even the matrix of morphism
(which has more information), as it can be seen from the following
example.

Let $\phi_1$ be as follows: $0 \to 01$, $1 \to 120$, $2 \to 2$, and
$\phi_2$ be as follows: $0 \to 01$, $1 \to 210$, $2 \to 2$. Then
these two morphisms have identical matrices of morphism, but
$\phi_1^\infty(0)$ is almost periodic, while $\phi_2^\infty(0)$ is
not. Indeed, in $\phi_2^\infty(0)$ there are arbitrary long segments
like 222\dots22, so $\phi_2^\infty(0) \notin \AP$. There is no such
problem in $\phi_1^\infty(0)$. Since 0 occurs in both $\phi_1(0)$
and $\phi_1(1)$, and $22$ does not occur in $\phi_1^\infty(0)$, it
follows that 0 occurs in $\phi_1^\infty(0)$ with bounded distances.
Thus $\phi_1^m(0)$ for every $m \geqslant 0$ occurs in
$\phi_1^\infty(0)$ with bounded distances, so $\phi_1^\infty(0) \in
\AP$. See Theorem~\ref{purenoneraseing-criterion} for a general
criterion of almost periodicity in the case of fixed points of
non-erasing morphisms.

To introduce a bit the notion of almost periodicity, let us
formulate an interesting result on this topic. It seems to be first
proved in~\cite{Cobh72}, but also follows from the results
of~\cite{Prit06c}. For $x\in A^\N$, $y\in B^\N$ define $x \times y
\in (A\times B)^\N$ such that $(x \times y)(i) = \langle
x(i),y(i)\rangle$.

\begin{proposition} \label{APtimesPer}
If $x$ is almost periodic and $y$ is periodic, then $x \times y$ is
almost periodic.
\end{proposition}

\section{Pure Morphic Sequences Generated by
Non-erasing Morphisms} \label{Pure}

Here we consider the case of morphic sequence of the form
$\phi^\infty(s)$ for non-erasing~$\phi$. We present an algorithm
that determines whether a morphic sequence $\phi^\infty(s)$ is
almost periodic given an alphabet $A$, a morphism $\phi$ and a
symbol $s \in A$.

Suppose we have $A$, $\phi$ and $s \in A$, such that $|A| = n$,
$\max_{b \in A} |\phi(b)| = k$, $s$ begins $\phi(s)$. Remember that
we suppose that all the symbols from $A$ appear in~$\phi^\infty(s)$.

Divide $A$ into two parts. Let $I$ be the set of all symbols $b \in
A$ such that $|\phi^m(b)| \to \infty$ as $m \to \infty$. Denote $F =
A \setminus I$, it is the set of all symbols $b$ such that
$|\phi^m(b)|$ is bounded. Also define $E \subseteq F$ to be the set
of all symbols $b$ such that $|\phi(b)| = 1$.

We can find a decomposition $A = I \sqcup F$ in $\poly(n,k)$-time as
follows.

Find $E$. Then find all the cycles in $G$ with all the vertices
lying in $E$. Join all the vertices of all these cycles in a set
$D$. This set is stabilizing: $F$ is the set of all vertices in $G$
such that all infinite paths starting from them stabilize in~$D$.
Polynomiality can be checked easily.

Construct ``a graph of left tails'' $L$ with marked edges. Its set
of vertices is~$I$. From each vertex $b$ exactly one edge goes off.
To construct this edge, find a representation $\phi(b) = uáv$, where
$c \in I$, $u$ is the maximal prefix of $\phi(b)$ containing only
symbols from $F$. It follows from the definitions of $I$ and $F$
that $u$ does not coincide with $\phi(b)$, that is why this
representation is correct. Then construct in $L$ an edge from $b$ to
$c$ and write $u$ on it.

Analogously we construct ``a graph of right tails'' $R$. (In this
case we consider representations $\phi(b) = váu$ where $u \in F^*$,
$c \in I$.)

Now we formulate a general criterion.

\begin{theorem} \label{purenoneraseing-criterion}
A sequence $\phi^\infty(s)$ is almost periodic iff \\
1) $G$ restricted to $I$ is strongly connected;\\
2) in graphs $L$ and $R$ on each edge of each cycle an empty word
$\Lambda$ is written.
\end{theorem}

It seems that full and detailed proof of this theorem can only
confuse a reader, rather than a proof sketch.

\begin{proof}[Proof sketch]
By Lemma~\ref{first-criterion} for almost periodicity it is
necessary and sufficient to check whether symbol $s$ occurs
infinitely many times with bounded distances.

For every symbol $b \in I$ the symbol $s$ should occur in some
$\phi^l(b)$, that is what the 1st part of the criterion says.

Furthermore, in the sequence $\phi^\infty(s)$ all the segments of
consecutive symbols from $F$ should be bounded. Indeed, every such
segment consists only of symbols from $F$, but $s \notin F$. That is
what the 2nd part of the criterion means, let us explain why.

Consider some $v = buc$ occurring somewhere in $\phi^\infty(a)$,
where $b,c \in I$, $u \in F^*$. Every element of sequence of words
$v, \phi(v), \phi^2(v), \phi^3(v), \dots$ occurs in
$\phi^\infty(s)$. Somewhere in the middle of $\phi^l(v) =
\phi^l(b)\phi^l(u)\phi^l(c)$ a word $\phi^l(u)$ occurs. As $l$
increases, some words from $F^*$ might stick to $\phi^l(u)$ from
left or right for these words can come from $\phi^l(b)$ or
$\phi^l(c)$. These words exactly correspond to those written on
edges of $L$ or $R$. The 2nd part of the criterion exactly says that
this situation can happen only finitely many times, until we get to
some cycle in $L$ or $R$.
\end{proof}

Let us consider examples with $\phi_1$ and $\phi_2$ from the end of
Section~\ref{Preliminaries}. In both cases $I = \{0, 1\}$, $F =
\{2\}$. On every edge of $R$ in both cases $\Lambda$ is written.
Almost the same is true for $L$: the only difference is about the
edge going from 1 to 1. In the case of $\phi_1$ an empty word is
written on this edge, while in the case of $\phi_2$ a word $2$ is
written. That is why $\phi_1^\infty(0)$ is almost periodic, while
$\phi_2^\infty(0)$ is not.

\begin{corollary}
If for all $b \in A$ we have $|\phi(b)| \geqslant 2$, then
$\phi^\infty(s)$ is almost periodic iff $\phi$ is primitive.
\end{corollary}
\begin{proof}
Follows from Theorem~\ref{purenoneraseing-criterion}. In that case
$A = I$, and on all the edges of $L$ and $R$ the empty word is
written.
\end{proof}

\begin{corollary}\label{algorithm-purenoneraseing-criterion}
There exists a $\poly(n,k)$-algorithm that says whether
$\phi^\infty(s)$ is almost periodic.
\end{corollary}
\begin{proof}
Conditions from Theorem~\ref{purenoneraseing-criterion} can be
checked in polynomial time.
\end{proof}

It also seems useful to formulate an explicit version of the
criterion for the binary case. We do it without any additional
assumptions, opposite to the previous.

\begin{corollary}\label{purenoneraseing-criterion-binary}
For non-erasing $\phi\colon \B \to \B$ that is prolongable on $0$ a
sequence $\phi^\infty(0)$ is almost periodic iff one of the
following conditions holds:\\
1) $\phi(0)$ contains only 0s;\\
2) $\phi(1)$ contains 0;\\
3) $\phi(1) = \Lambda$;\\
4) $\phi(1) = 1$ and $\phi(0) = 0u0$ for some word $u$.
\end{corollary}

\section{Uniform Morphisms} \label{Uniform}

Now we deal with morphic sequences obtained by uniform morphisms.
Again we present a polynomial-time algorithm for solving the problem
in this situation.

Suppose we have an alphabet $A$, a morphism $\phi\colon A^* \to
A^*$, a coding $h\colon A \to B$, and $s \in A$, such that $|A| =
n$, $|B| \leqslant n$, $\forall b \in A\ |\phi(b)| = k$, $s$ begins
$\phi(s)$. We are interested in whether $h(\phi^\infty(s))$ is
almost periodic. Sequences of the form $h(\phi^\infty(s))$ with
$\phi$ being $k$-uniform are also called $k$-automatic
(see~\cite{AlShall03}).

\subsection{Equivalence Relations and Uniform Morphisms}\label{somedef}

For each $l \in \N$ define an equivalence relation on $A$: $b \sim_l
c$ iff $h(\phi^l(b)) = h(\phi^l(b))$. We can easily continue this
relation on $A^*$: $u \sim_l v$ iff $h(\phi^l(u)) = h(\phi^l(v))$.
In fact, this means $|u| = |v|$ and $u(i) \sim_l v(i)$ for all $i$,
$1 \leqslant i \leqslant |u|$.

Let $B_m$ be the Bell number, i.~e., the number of all possible
equivalence relations on a finite set with exactly $m$ elements,
see~\cite{Weis}. As it follows from this article, we can
estimate~$B_m$ in the following way.

\begin{lemma}\label{equivalence-bound}
$2^m \leqslant B_m \leqslant 2^{Cm\log m}$ for some constant $C$.
\end{lemma}

Thus the number of all possible relations $\sim_l$ is not greater
than $B_n = 2^{O(n\log n)}$. Moreover, the following lemma gives a
simple description for the behavior of these relations as $l$ tends
to infinity.

\begin{lemma}
If $\sim_r$ equals $\sim_s$, then $\sim_{r+p}$ equals $\sim_{s+p}$
for all $p$.
\end{lemma}
\begin{proof}
Indeed, suppose $\sim_r$ equals $\sim_s$. Then $b \sim_{r+1} c$ iff
$\phi(b) \sim_r \phi(c)$ iff $\phi(b) \sim_s \phi(c)$ iff $b
\sim_{s+1} c$. So if $\sim_r$ equals $\sim_s$, then $\sim_{r+1}$
equals $\sim_{s+1}$, which implies the lemma statement.
\end{proof}

This lemma means that the sequence $(\sim_l)_{l \in \N}$ turns out
to be ultimately periodic with a period and a preperiod both not
greater than $B_n$. Thus we obtain the following

\begin{lemma}\label{equivalence-periodicity}
For some $p,q \leqslant B_n$ we have for all $i$ and all $t > p$
that $\sim_{t}$ equals~$\sim_{t + iq}$.
\end{lemma}

\subsection{Criterion}\label{tocriterion}

Now we are trying to get a criterion which we could check in
polynomial time. Notice that the situation is much more difficult
than in the pure case because of a coding allowed. In particular,
the analogue of Lemma~\ref{first-criterion} for non-pure case does
not hold.

We will move step by step to the appropriate version of the
criterion reformulating it several times.

This proposition is quite obvious and follows directly from the
definition of almost periodicity since all $h(\phi^m(a))$ are the
prefixes of $h(\phi^\infty(a))$.

\begin{proposition}
A sequence $h(\phi^\infty(s))$ is almost periodic iff for all $m$
the word $h(\phi^m(s))$ occurs in $h(\phi^\infty(s))$ infinitely
often with bounded distances.
\end{proposition}

And now a bit more complicated version.

\begin{proposition}\label{second-uniform-criterion}
A sequence $h(\phi^\infty(s))$ is almost periodic iff for all $m$
the symbols that are $\sim_m$-equivalent to $s$ occur in
$\phi^\infty(s)$ infinitely often with bounded distances.
\end{proposition}
\begin{proof}
$\Leftarrow$. If the distance between two consecutive occurrences in
$\phi^\infty(s)$ of symbols that are $\sim_m$-equivalent to $s$ is
not greater than $t$, then the distance between two consecutive
occurrences of $h(\phi^m(s))$ in $h(\phi^\infty(s))$ is not greater
than $tk^m$.

$\Rightarrow$. Suppose $h(\phi^\infty(s))$ is almost periodic. Let
$y_m = 0 1 2\dots(k^m-2)(k^m-1) 0 1 \dots (k^m-1) 0 \dots$ be a
periodic sequence with a period $k^m$. Then by
Proposition~\ref{APtimesPer} a sequence $h(\phi^\infty(s)) \times
y_m$ is almost periodic, which means that the distances between
consecutive $k^m$-aligned occurrences of $h(\phi^m(s))$ in
$h(\phi^\infty(s))$ are bounded. It only remains to notice that if
$h(\phi^\infty(s))[ik^m, (i+1)k^m-1] = h(\phi^m(s))$, then
$\phi^\infty(s)(i) \sim_m s$.
\end{proof}

Let $Y_m$ be the following statement: symbols that are
$\sim_m$-equivalent to $s$ occur in $\phi^\infty(s)$ infinitely
often with bounded distances.

Suppose for some $T$ that $Y_T$ is true. This implies that
$h(\phi^T(s))$ occurs in $h(\phi^\infty(s))$ with bounded distances.
Therefore for all $m \leqslant T$ a word $h(\phi^m(s))$ occurs in
$h(\phi^\infty(s))$ with bounded distances since $h(\phi^m(s))$ is a
prefix of $h(\phi^T(s))$. Thus we do not need to check the
statements $Y_m$ for all $m$, but only for all $m \geqslant T$ for
some~$T$.

Furthermore, it follows from Lemma~\ref{equivalence-periodicity},
that we are sufficient to check the only one such statement as in
the following

\begin{proposition}
For all $r \geqslant B_n$: a sequence $h(\phi^\infty(s))$ is almost
periodic iff the symbols that are $\sim_r$-equivalent to $s$ occur
in $\phi^\infty(s)$ infinitely often with bounded distances.
\end{proposition}

And now the final version of our criterion.

\begin{proposition}\label{last-uniform-criterion}
For all $r \geqslant B_n$: a sequence $h(\phi^\infty(s))$ is almost
periodic iff for some $m$ the symbols that are $\sim_r$-equivalent
to $s$ occur in $\phi^m(b)$ for all $b \in A$.
\end{proposition}

Indeed, if the symbols of some set occur with bounded distances,
then they occur on each $k^m$-aligned segment for some sufficiently
large~$m$.

\subsection{Polynomiality}\label{polynomiality}

Now we explain how to check a condition from
Proposition~\ref{last-uniform-criterion} in polynomial time. We need
to show two things: first, how to choose some $r \geqslant B_n$ and
to find in polynomial time the set of all symbols that are
$\sim_r$-equivalent to $s$ (and this is a complicated thing keeping
in mind that $B_n$ is exponential), and second, how to check whether
for some $m$ the symbols from this set for all $b \in A$ occur
in~$\phi^m(b)$.

Let us start from the second. Suppose we have found the set $H$ of
all the symbols that are $\sim_r$-equivalent to $s$. For $m \in \N$
let us denote by $P_m^{(b)}$ the set of all the symbols that occur
in $\phi^m(b)$. Our aim is to check whether exists $m$ such that for
all $b$ we have $P_m^{(b)} \cap H \ne \varnothing$. First of all,
notice that if $\forall b\ P_m^{(b)} \cap H \ne \varnothing$, then
$\forall b\ P_l^{(b)} \cap H \ne \varnothing$ for all $l \geqslant
m$. Second, notice that the sequence of tuples of sets
$((P_m^{(b)})_{b \in \Sigma})_{m=0}^\infty$ is ultimately periodic.
Indeed, the sequence $(P_m^{(b)})_{m=0}^\infty$ is obviously
ultimately periodic with both period and preperiod not greater
than~$2^n$ (recall that $n$ is the size of the alphabet $\Sigma$).
Thus the period of $((P_m^{(b)})_{b \in \Sigma})_{m=0}^\infty$ is
not greater than the least common divisor of that for
$(P_m^{(b)})_{m=0}^\infty$, $b \in A$, and the preperiod is not
greater than the maximal that of $(P_m^{(b)})_{m=0}^\infty$. So the
period is not greater than $(2^n)^n = 2^{n^2}$ and the preperiod is
not greater than~$2^n$. Third, notice that there is a
polynomial-time-procedure that given a graph corresponding to some
morphism $\psi$ (see Section~\ref{Preliminaries} to recall what is
the graph corresponding to a morphism) outputs a graph corresponding
to morphism~$\psi^2$. Thus after repeating this procedure $n^2 + 1$
times we obtain a graph by which we can easily find $(P_{2^{n^2} +
2^n}^{(b)})_{b \in \Sigma}$, since $2^{n^2 + 1} > 2^{n^2} + 2^n$.

Similar arguments, even described with more details, are used in
deciding our next problem. Here we present a polynomial-time
algorithm that finds the set of all symbols that are
$\sim_r$-equivalent to $s$ for some $r \geqslant B_n$.

We recursively construct a series of graphs $T_i$. Let its common
set of vertices be the set of all unordered pairs $(b,c)$ such that
$b,c \in A$ and $b \ne c$. Thus the number of vertices is
$\frac{n(n-1)}2$. The set of all vertices connected with $(b,c)$ in
the graph $T_i$ we denote by $V_i(b,c)$.

Define a graph $T_0$. Let $V_0(b,c)$ be the set $\{(\phi(b)(j),
\phi(c)(j)) \mid j = 1, \ldots, k, \phi(b)(j) \ne \phi(c)(j)\}$. In
other words, $b \sim_{l+1} c$ if and only if $x \sim_l y$ for all
$(x,y) \in V_0(b,c)$.

Thus $b \sim_2 c$ if and only if for all $(x,y) \in V_0(b,c)$ for
all $(z,t) \in V_0(x,y)$ we have $z \sim_0 t$. For the graph $T_1$
let $V_1(b,c)$ be the set of all $(x,y)$ such that there is a path
of length 2 from $(b,c)$ to $(x,y)$ in $T_0$. The graph $T_1$ has
the following property: $b \sim_2 c$ if and only if $x \sim_0 y$ for
all $(x,y) \in V_1(b,c)$. And even more generally: $b \sim_{l+2} c$
if and only if $x \sim_l y$ for all $(x,y) \in V_1(b,c)$.

Now we can repeat operation made with $T_0$ to obtain $T_1$. Namely,
in $T_2$ let $V_2(b,c)$ be the set of all $(x,y)$ such that there is
a path of length 2 from $(b,c)$ to $(x,y)$ in $T_1$. Then we obtain:
$b \sim_{l+4} c$ if and only if $x \sim_l y$ for all $(x,y) \in
V_2(b,c)$.

It follows from Lemma~\ref{equivalence-bound} that $\log_2 B_n
\leqslant Cn\log n$. Thus after we repeat our procedure $r = [Cn\log
n]$ times, we will obtain the graph $T_l$ such that $b \sim_{2^r} c$
if and only if $x \sim_0 y$ for all $(x,y) \in V_2(b,c)$. Recall
that $x \sim_0 y$ means $h(x) = h(y)$, so now we can easily compute
the set of symbols that are $\sim_{2^r}$-equivalent to~$s$.

\section{Monadic Theories}
\label{Monadic}

Combinatorics on words is closely connected with the theory of
second order monadic logics. Here we just want to show some examples
of these connections. More details can be found, e.~g.,
in~\cite{Sem79,Sem83}.

We consider monadic logics on $\N$ with the relation ``$<$'', that
is, first-order logics where also unary finite-value function
variables and quantifiers over them are allowed. We also suppose
that we know some fixed finite-value function $x\colon \N \to
\Sigma$ and can use it in our formulas. Such a theory is denoted by
$\MT x$ and is called \emph{monadic theory of~$x$}.

The main question here can be the question of decidability, that is,
does there exist an algorithm that given a sentence in a theory says
whether this sentence is true of false.

The criterion of decidability for monadic theories of almost
periodic sequences can be formulated in terms of some their very
natural characteristic, namely, almost periodicity regulator. An
almost periodicity regulator of an almost periodic sequence $x$ is a
function $f\colon \N \to \N$ such that every factor $u$ of $x$ of
length $n$ occurs in each factor of $x$ of length $f(n)$. So an
almost periodicity regulator somehow regulates how periodic a
sequence is. Notice that an almost periodicity regulator of a
sequence is not unique: every function greater than regulator is
also a regulator.

\begin{theorem}[Semenov 1983 \cite{Sem83}] \label{Semenov-monadic}
If $x$ is almost periodic, then $\MT x$ is decidable iff $x$ and
some its almost periodicity regulator are computable.
\end{theorem}

The following result was obtained recently, but uses the technics
already used in \cite{Sem79,Sem83}.

\begin{theorem}[Carton, Thomas 2002 \cite{CarThom02}] \label{Thomas-monadic}
If $x$ is morphic, then $\MT x$ is decidable.
\end{theorem}

A curious result can be implied from two these theorems.

\begin{corollary} \label{ap-morphic-regulator-corollary}
If $x$ is both morphic and almost periodic, then some its regulator
is computable.
\end{corollary}
\begin{proof}
Indeed, if $x$ is morphic, then by Theorem~\ref{Thomas-monadic} the
theory $\MT x$ is decidable. Since $x$ is almost periodic, from
Theorem~\ref{Semenov-monadic} it follows that some almost
periodicity regulator of $x$ is computable.
\end{proof}

Notice that Corollary~\ref{ap-morphic-regulator-corollary} does not
imply the existence of an algorithm that given a morphic sequence
computes some almost periodicity regulator of this sequence whenever
it is almost periodic (but probably this algorithm can be
constructed after deep analyzing the proofs of
Theorems~\ref{Semenov-monadic} and~\ref{Thomas-monadic} and showing
uniformity in a sense). And it also does not imply the decidability
of almost periodicity for morphic sequences. This decidability also
does not imply Corollary~\ref{ap-morphic-regulator-corollary}.

By the way, Corollary~\ref{ap-morphic-regulator-corollary} allows us
to hope that these algorithms exist. Though the formulation of this
statement uses only combinatorics on words, the proof also involves
the theory of monadic logics. Of course, it would be interesting to
find a simple combinatorial proof of the result.

And the last remark here is that
Corollary~\ref{ap-morphic-regulator-corollary} (and its probable
uniform version) seems to be the best progress that we can obtain by
this monadic approach. One could try to express in the monadic
theory of morphic sequence (which is decidable by
Theorem~\ref{Thomas-monadic}) the property of almost periodicity,
but it turns out to be impossible.

\section{In General Case} \label{General}

We have described two polynomial-time algorithms, but without any
precise bound for their working time. Of course, it can be done
after deep analyzing of all the previous, but is probably not so
interesting.

It is not still known whether the problem of determining almost
periodicity of arbitrary morphic sequence is decidable.
Corollary~\ref{ap-morphic-regulator-corollary} somehow supports the
conjecture of decidability (but even does not follow from this
conjecture!).

Theorem 7.5.1 from \cite{AlShall03} allows us to represent an
arbitrary morphic sequence $h(\phi^\infty(s))$ as
$g(\psi^\infty(b))$ where $\psi$ is non-erasing. So it is sufficient
to solve our main problem for $h(\phi^\infty(s))$ with non-erasing
$\phi$.

It seems that the general problem is tightly connected with a
particular case of $h(\phi^\infty(a))$ where $|\phi(b)| \geqslant 2$
for each $b \in A$. There is no strict reduction to this case but
solving problem in this case can help to deal with general
situation.

The problem of finding an effective periodicity criterion in the
case of arbitrary morphic sequences is also of great interest, as
well as criteria for variations with periodicity and almost
periodicity: ultimate periodicity, generalized almost periodicity,
ultimate almost periodicity (see~\cite{Prit06c} for definitions). If
one notion is a particular case of another, it does not mean that
corresponding criterion for the first case is more difficult (or
less difficult) than for the second.

\subsection*{Acknowledgements}

The author is grateful to \fbox{An.~Muchnik} and A.~Semenov for
their permanent help in the work, to A.~Frid, M.~Raskin, K.~Saari
and to all the participants of Kolmogorov seminar, Moscow
\cite{KolmSem}, for fruitful discussions, and also to anonymous
referees for very useful comments.

\end{document}